\documentclass{article}
\usepackage[utf8]{inputenc}

\usepackage{graphicx}
\usepackage[]{caption}

\usepackage[shortlabels]{enumitem}
\setlist[enumerate, 1]{1)}

\usepackage{amsmath}

\usepackage{array}
\usepackage{epstopdf}

\usepackage{amsfonts}

\usepackage{authblk}

\newcolumntype{P}[1]{>{\centering\arraybackslash}p{#1}}

\begin{document}

\title{Common Foundations of Optimal Control Across the Sciences: evidence of a free lunch}

\author[1]{Benjamin Russell, Herschel Rabitz}

\affil[1]{Department of Chemistry, Princeton University, Princeton, NJ 08540, USA}

\maketitle

\begin{abstract}
A common goal in the sciences is optimization of an objective function by selecting control variables such that a desired outcome is achieved.
This scenario can be expressed in terms of a control landscape of an objective considered as a function of the control variables.
At the most basic level, it is known that the vast majority of quantum control landscapes possess no traps, whose presence would hinder reaching the objective.
This paper reviews and extends the quantum control landscape assessment, presenting evidence that the same highly favorable landscape features exist in many other domains of science.
The implications of this broader evidence are discussed.
Specifically, control landscape examples from quantum mechanics, chemistry, and evolutionary biology are presented.
Despite the obvious differences, commonalities between these areas are highlighted within a unified mathematical framework.
This mathematical framework is driven by the wide ranging experimental evidence on the ease of finding optimal controls (in terms of the required algorithmic search effort beyond laboratory set up overhead).
The full scope and implications of this observed common control behavior pose an open question for assessment in further work.
\end{abstract}

\section{Introduction to Control Landscapes}

Cybernetics is often referred to as ``the science of communications and automatic control systems in both machines and living things'' \cite{wiener}.
This paper, based on a presentation at the IEEE 2016 conference on Norbert Wiener, fully embraces the spirit of cybernetics by considering control throughout the sciences and beyond in the $21^{\text{st}}$ century.

Since Weiner wrote his original work, optimization has become ubiquitous throughout the sciences and engineering.
In the laboratory and in many industrial settings, one seeks to optimally control a physical, chemical or biological system to induce a specific transformation using a adequate resource set.
In natural evolution it is generally accepted that Nature is performing a stochastic optimization expressed in the survival of the fittest.
Although not humanly driven, we also discuss evolution in a control context.

A control landscape is an objective function, or \emph{goal}, subject to optimization by choosing control parameters in a system.
The landscape metaphor is established by considering the graph, or higher dimensional analogue, of the objective as a function of the control parameters.
The concept of a fitness landscape in evolutionary biology was introduced \cite{write}, where it is prevalent.
In recent years, the study of control landscapes has focused extensively on quantum systems which evolve according to Schr\"odinger's equation within a well defined mathematical structure.
As such, this work will primarily consider the inherent simplicity of control landscapes of quantum systems and further present evidence that these features exist also outside of quantum mechanics and propose an underlying mathematical basis for this commonality.
In particular, evidence abounds in chemistry and material science, evolution  and engineering that control landscape behavior exists like that found in quantum mechanics.

Control systems come in many forms, however, there are generic mathematical structures underlying many practical cases.
In order to set the framework for the remainder of this paper, we first describe a general class of mathematical problems given by a system of differential equations comprising a physical model of the system under control:
\begin{align}
	\label{feqn}
	\frac{d x(t)}{dt} = F(x(t), w(t))
\end{align}
where $x(t)$ is the state vector at time $t$, and $w(t)$ is the control chosen by an experimenter or by Nature as in biological evolution.
This class of equations is vast and includes diverse examples from many domains.
The Schr\"odinger equation is of this form, as are some models used in molecular dynamics and evolutionary biology \cite{jones}.
In some cases the variables defining the structural form of $F$ may also be considered part of the control.

In order to inquire about \emph{optimal} control, one must have an objective function to optimize.
This function is often represented by: $\mathcal{F}[w] = J(x(T))$, a function of the state at some given final time $T$, which we seek to maximize by choosing (possibly time dependent) control variables, or simply \emph{controls}, represented by $w$.
Such functions are known as terminal cost/payoff functions, as they only depend on the state at some final time without additional `run-time' costs depending on $w(t)$.

In such scenarios it becomes prescient to ask about the structure of the control landscape, which is crucial in establishing the feasibility of finding optimal controls.
We will render this feasibility assessment in terms of two issues:
\begin{enumerate}
\item $\label{is1}\text{The existence of Landscape traps}$ 
\item $\label{is2}\text{Landscape optimization convergence rate}$
\end{enumerate}
Both of the issues \ref{is1}, \ref{is2} relate strongly to the nature of the algorithms seeking an optimal control.
The presence of landscape traps, that is sub-optimal extrema, would call for a stochastic algorithm to `step over' such features.
The second issue \ref{is2} addresses the search effort required to locate an optimal control.
In laboratory cases where there can be hundreds of control variables present, out to biological evolution for which there are approximately $10^9$ gnomic sites for Nature to select nucleic acids, the success of a multitude of experimental and natural optimization processes indicates that the structure of these landscapes is favorable (i.e. free from traps under reasonable assumptions) permitting rapid convergence despite the curse of dimensionality.

\begin{figure}[h!]
\centering
	\includegraphics[width=0.4\textwidth]{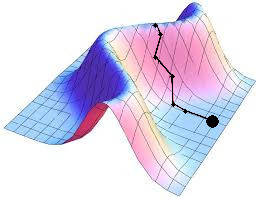}
  \caption{A control landscape with a plateau (showing a one dimensional critical manifold of global optima) but no with traps. Landscapes of this type, or even with additional saddle features, are highly favorable for finding optimal controls.}
\label{fig:cl1}
\end{figure}

\begin{figure}[h!]
\centering
	\includegraphics[width=0.6\textwidth]{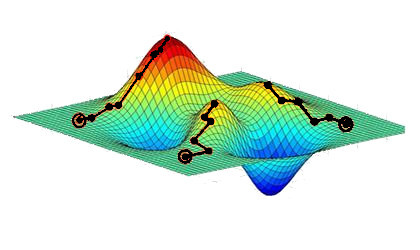}
  \caption{A local gradient search climbing the landscape either reaching dead-end sub-optimal traps or the desired true global optimum, depending on the starting point. Landscapes of this type, especially in high dimensions, are highly problematic for finding optimal controls.}
\label{fig:cl2}
\end{figure}

In principle, and perhaps intuitively, one might expect the complexity of many control systems to result in landscapes that possess large numbers of local optima.
We shall however argue, that a specific notion of complexity is favorable for control optimization rather than deleterious, to actually reduce the possibility of traps.
Figures (\ref{fig:cl1}) and (\ref{fig:cl2}) illustrate some low dimensional (two control parameters) examples.
In practice typically, many more control parameters than two are present and control landscapes cannot be directly visualized.

The \emph{critical point topology} of a control landscape determines the behavior of local optimization algorithms, such as gradient ascent and even non-local stochastic algorithms could be greatly hindered by a rough, high dimensional landscape.
An understanding of the critical point topology, i.e. the set of controls where $\nabla \mathcal{F} = 0$, permits assessing the ease of finding optimal controls.
Saddle points could exist on some landscapes, but they would only slow down a search rather than stop it from proceeding to full optimization.
For an application of a gradient method in laboratory practice see \cite{labgrad} in which the algorithm is applied to spectrally filtered and integrated second harmonic generation as well as excitation of atomic rubidium.
In this work, monotonic convergence to maximum fidelity is seen within practical laboratory timescales.
Work on assessing the experimental relevance of saddle points has also been undertaken \cite{qqcon2} in which their impact was found to be negligible, as theoretically expected.

The remainder of this paper is organized as follows.
Section \ref{evid} summarizes the experimental and simulation behavior seen when seeking optimal controls in the domain of quantum phenomena, chemistry and material science, and natural evolution.
Importantly, this evidence stands as a body of empirical facts on the favorable structure of landscapes that demands an explanation, which we claim has a common foundation.
Section \ref{gener} focuses on that explanation in quantum control, wherein detailed and exacting mathematical models exist.
Section \ref{beyo} starts with a brief summary of the situation in chemistry and material science, as well as evolution and the reader will be referred to the literature for specific details of these analyses.
The most challenging case of equation \ref{feqn} covers such a wide range of behavior that achieving a full landscape analysis has not yet been achieved.
But, we lay out the framework of conditions for finding favorable landscapes in this broad class of systems and offer arguments that the conditions plausibly hold.
In this regard, we further offer a well defined mathematical conjecture to be investigated in the future.
The evidence in section \ref{evid} speaks to the validity of the this conjecture across varied domains, and if further analysis expands the present support, it would offer a highly unified picture of optimization across the sciences and engineering.

\section{Evidence of Common Landscape Structure in the Sciences}

\label{evid}

Recent work \cite{controlvast} has identified evidence of common landscape structure from the control of chemical, physical and biological evolutionary processes.
Both the success of the optimization throughout these diverse areas, and the remarkably efficient search effort required is summarized in table \ref{tabsum}, alongside the nature of the sufficient conditions required to theoretically establish this simplicity.
\begin{table}[h!]
\begin{tabular}{P{2.5cm} || P{3cm} | P{3cm} | P{3cm}}
 & Quantum Control & Chemistry and Material Science & Natural Evolution \\ \hline \hline 
Evidence implying simplicity or demonstrating a trap free landscape & Extensive successful numerical simulations and some experimentally observed landscapes & Many successful experiments, including direct landscape observations & Many diverse perturb and observe experiments free from evident traps, and the existence of complex life forms \\ \hline
Sufficient assumptions for proving the structure of a trap free landscape & Three precise physical assumptions & The same assumptions as in quantum mechanics, but also the ability to manipulate the environment & Assumption that all probability distributions over a species population can be created by appropriate genomic variations
\end{tabular}
\caption{Summary of landscape analysis evidence and nature of the associated sufficient conditions in different areas. Importantly, the sufficient conditions which support the existence of trap free landscapes in all three categories are essentially the same, although the analysis methodology is different in each domain.}
\label{tabsum}
\end{table}

\newpage

\subsection{Quantum control}

\begin{figure}[h!]
\centering
	\includegraphics[width=0.4\textwidth]{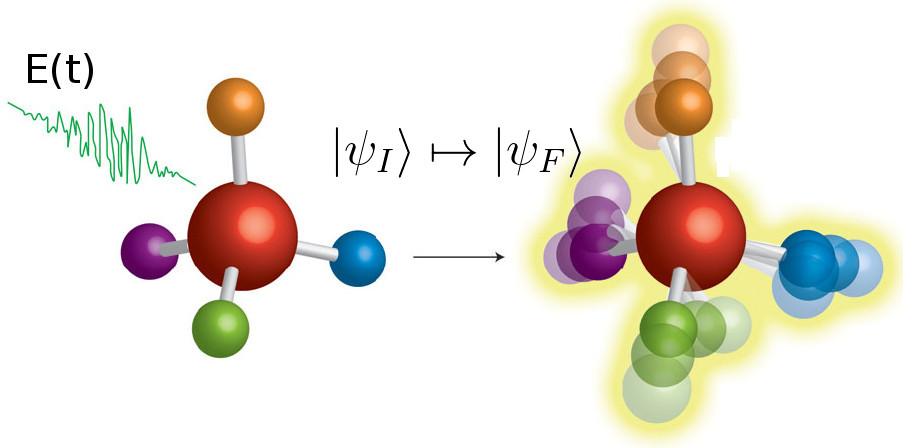}
  \caption{A molecule under vibrational control by a laser field $E(t)$.}
\label{zapmol}
\end{figure}

Control of a quantum system using electromagnetic fields generally encompasses one of the following tasks:
\begin{enumerate}
\item Control the wave function (or density matrix $\rho$), as illustrated in figure \ref{zapmol}, of a quantum system to create a specific state transformation $|\psi_{\text{I}} \rangle \mapsto |\psi_{\text{F}}\rangle$ \cite{wfctl, Wein}
\item Maximize the expectation of a particular quantum observable $\hat{O}$ expectation value \cite{qopcon}
\item Create a specific quantum gate $G$, which is a unitary transformation \cite{qconlan}
\end{enumerate}

There is strong numerical \cite{riv, evd1}, mathematical \cite{qclrev, qcrev2} and some experimental evidence \cite{qqcon1, qqcon2} that the landscapes of closed, finite dimensional quantum systems are trap free, corroborating the analytic claims when the assumptions are met.
A few cases deliberately constructed to possess traps \cite{qclcl, conqcl, axis} are however known.
Work in \cite{riv} assesses the few known quantum control examples with traps to better understand the volume of the traps.
Analysis in \cite{wusing} goes on further to numerically search for and analyze the typical singularities in quantum control landscapes for four level systems, and concludes that all the identified singularities are saddles.
The work in \cite{mepol} studies the effect upon the landscape of terms beyond linear order in the control within the Hamiltonian.
An additional term, quadratic in the control field, is added to the Hamiltonian.
It is found that the effect on the control landscape involves the removal of some specific traps known to exist without such a term.
This finding is a step towards a general conjecture in section \ref{beyo} that nominal system complexity aids the ability to find optimal controls.

In this work we will focus on closed quantum systems, i.e., ones which don't interact significantly with their environment.
However, there is existing work on the landscapes for observable preparation in open quantum systems \cite{openland}, and a proof that any two level open system (meeting simple further assumptions) is free from traps \cite{twoopen}.

\subsection{Chemical yield and material property optimization}

Evidence relevant to the quality of discovered, and the optimization search effort required to find optimal controls, abound in chemical and material sciences \cite{controlvast, optichemtab}.
These data are from several different objectives corresponding to common goals in chemistry \cite{optichemeasy,optichemtab} of chemical yield and material property optimization.
A clear pattern is present, especially within robotized experiments, that the number of samples required to obtain effective controls is typically infinitesimal compared to the full search space of chemical and operating conditions.
This behavior strongly suggests that the processes being optimized possess landscapes rife with many global optima, with few or no traps thereby permitting highly rapid ascent to a global optimum in control space.

\subsection{Optimization in biological evolution}

The concept of a fitness landscape is pervasive in natural evolution.
In this context, the fitness landscape in genetics can be thought of as a graph of genotypic allele frequency in a population, against phenotype fitness \cite{wright}, which is further developed in \cite{optevo}.
In this sense, the genes of a species population (or even over multi-species populations) are the `controls', and the fitness is the optimality condition.
However, many subtle variants exist in the exact nature of the fitness studied including using population growth rate as a measure of fitness, or average number of individual offspring.

One important means of assessing genetic landscapes for traps in the laboratory are `disturb and observe' experiments.
In one form of such experiments, a small change is made to the genome of an organism with a short life cycle, such as drosofila fruit flies.
Then the population of such genetically-`disturbed' organisms are allowed to freely evolve and the average population fitness is monitored to assess if it returns to the original, undisturbed value.
If the fitness returns to its original value, this is evidence that the fitness landscape, at least in the vicinity of the initial genomic distribution, is free from local optima.
In many known cases \cite{optevo}, the fitness was seen to return to its original value but with a different genome indicative of a level set at the top of the landscape, as shown in figure \ref{fig:cl2}.
Beyond such laboratory experiments lies the most stark evidence, that complex organisms exist.
The period of time taken for the appearance of bacteria, and beyond to higher forms of life, as only a few billion years, is very short given their inherent complexity and the required multitude of elaborate evolutionary steps.
These findings, particularly the latter, have left a puzzle especially considering that intuition would suggest that evolutionary landscapes with approximately $10^9$ nucleic acid sites would be very rugged.

The evidence in the scientific domains of quantum mechanics, chemistry and material science, and evolution exist in a vast body of literature, which must be taken as foundational facts to be reconciled, perhaps individually in each domain, or more enticingly in a sweeping consistent fashion as indicated in the second portion of table \ref{tabsum} and the remainder of this paper.
The details in treating the evidence differ in swinging between domains, but the analyses are similar, which eventually leads to conjecturing that the scope of observed landscape behavior extends to the general structure of equation \ref{feqn} upon satisfaction of three assumptions, enumerated in the remainder of the paper.

\section{Specification of Landscape Assumptions for Quantum Control}

\label{gener}

As the above evidence points towards the existence of common features in the optimal control of systems arising in diverse areas of science, this finding provides an enticing incentive to seek a unifying explanation for the origin of this simplicity.
The focus of the analysis in the literature thus far has been in quantum control, which, of all the domains, has the most detailed mathematical foundation to rest on.
This foundation has aided in assessing the way in which the key assumptions (i.e., sufficient conditions), and the conclusions they imply about trap free control landscapes, generalize to systems outside of quantum mechanics.

\label{threeass}

In the work \cite{rabscience}, the first essential principles of the analysis of quantum control landscapes were set out, as were a nascent version of the three assumptions sufficient for trap free quantum control landscapes which are summarized below, and later generalized in section \ref{beyo}.
Shortly after the introduction of the `photonic reagent' control concept \cite{reagents}, the search effort required for the discovery of effective quantum controls was quickly identified to be dramatically less than expected, both in simulation and mounting numbers of experiments.

While the case of a quantum system controlling another \cite{qqconreal} does fall within the class of non-linear control problems studied in this work (\ref{feqn}), here we restrict our attention to quantum systems under control by semi-classical fields, as is typical throughout the vast majority of the quantum control literature.
Closed systems evolve in time according to the Schr\"odinger equation:
\begin{align}
\label{scheqn}
\frac{d U_t}{dt} = iH[w(t)] U_t
\end{align}
which is an example system of \ref{feqn} for which the propagator $U_t$ evolves on the Lie group of $n \times n$ unitary matrices $U(n)$.
Here the Hamiltonian $H[w(t)]$ depends on the time dependent control function $w(t)$, which often represents an external laser field (i.e., a photonic reagent).
A common approximation in physical scenarios where the quantum state of a molecule is controlled by a laser is the dipole approximation:
\begin{align}
\label{dipole}
H[w(t)] = H_0 + w(t)H_c
\end{align}
For dipole systems of the form in \ref{dipole} (which includes the control of nuclear spins in NMR), an in depth landscape analysis has been performed for a control which implements a desired unitary transformation \cite{qclanalysisII}, quantum state transfer \cite{qclanalysisI} or maximizes a desired quantum observable \cite{qclanalysisop}.
The satisfaction of a set of three simple assumptions are known to guarantee the control landscape in the dipole approximation is free from traps in all these cases.

In the discussion of quantum control in this work, the cost function,
\begin{align}
\label{qcost}
J(U) :=  \left| \text{Tr}(G^{\dagger} U) \right|^2,
\end{align}
will be used to measure how well the unitary goal gate $G$ was implemented by a system for which the end-time propagator is $U_T \equiv U$.
Given a system where the assumptions hold, there may still be sub-optimal critical points introduced into the landscape by the cost function $J$ itself \ref{qcost}, however these are known analytically to all be `benign' saddles \cite{dom2} rather than true local optima.

The three assumptions of this analysis are described below and a proof that they are sufficient for a trap free landscape is included in Appendix \ref{trapapp}.

\subsection{Assumption I: the system is controllable}

\label{ass1}

The definition of controlability is that the final time $T$ and Hamiltonian $H$ are such that for every objective evolution of the system, there exists at least one control $w$ such that $U_T = G$.
For systems in the dipole approximation \ref{dipole}, a criterion ensuring controlability exists \cite{jurdjevic, altifini}.
This criterion, known as the \emph{Lie Algebra Rank Condition}, is necessary and sufficient for controllability.
It states that $H_0$ and $H_c$ must `generate' the whole algebra of observable (Hermitian matrices), in the sense that a basis can be created from a succession of Lie bracket expressions of the two generators, $\{H_0, H_c, [H_0, H_c], [[H_0, H_c], H_0] \ldots \}$, in order for the system to be controllable.
For a visual representation of checking this criterion see \cite{censpin}.
It is further known, that for typical quantum systems of the form \ref{dipole} the criterion holds \cite{altifini}.
Specifically, only a null set of Hamiltonians (a set of zero measure, or equivalently probability zero, if cases are chosen at random) fail to be controllable.

\subsection{Assumption II: The system is locally controllable in the direction of cost function gradient}

Local controllability is defined as the ability, given any control, to arbitrarily vary the systems state at the final time by varying the control infinitesimally.
The required landscape condition is a special case of this property, as we only require that the state at the final time can be varied in \emph{one} specific direction.
More explicitly, the second assumption is: for all controls $w(t)$ there exists a small change in the control $\delta w$ such that $\delta U_T$ has some component in the direction $\nabla J$.
That is,
\begin{align}
\langle \nabla J \big|_{U_T}, \delta U_T \rangle \neq 0, \forall \delta w.
\end{align}
This expression means that the control can be varied slightly to increase the objective function $J$ by `steering' $U_T$ in the direction $\nabla J$ and thus towards the goal (i.e. in the direction of increasing $J$).
Informally, this criterion is simply that it is possible to `steer' the system up the landscape by a small variation of the control.
This property can also be proven to typically hold, even well beyond systems in the dipole approximation.
Specifically, it has been shown that, in analogy with the results of \cite{altifini}, this second assumption only fails for a null set of Hamiltonians.
For a detailed discussion of the somewhat technical underlying mathematics of this result, which is based on an application of the parametric transversality theorem \cite{Thom}, see \cite{mesing}.

\subsection{Assumption III: The control resources are unrestricted}

This assumption means that the control $w$ is not limited in its form.
However, in practice there are always restrictions on the control in the laboratory.
Although this assumption technically needs to be adopted in order to mathematically prove that quantum control landscapes are almost all trap free, in practice adequate of freedom in the control often suffices.
Pragmatically, this reduces to allowing the control to have \emph{sufficient} freedom to exploit assumptions I and II \cite{searchcom}.
For a review of quantum control landscapes in the presence of severe constraints, see \cite{exconqcl, topqcl, dono, qclcons2, searchsev}.
A commonly applied constraint is to limit the total field fluence: $Q=\int_{0}^{T} |E(t)|^2 dt$; this constraint is known to be able to introduce traps into the landscape above a certain critical value.
We generally expect that traps will appear in quantum control landscapes when severe constraints are imposed upon the control fields.
However, the exact conditions when constraints introduce traps into the landscape is not known analytically and could form the basis of future work as it is a topic with practical significance.

Reiterating the main conclusion, upon satisfaction of all the assumptions about quantum control it is known that be almost all landscapes are free from traps.

\section{Landscapes Beyond Quantum Mechanics}

\label{beyo}

Building on the previous results, this section draws a general conclusion and conjectures about about the behavior of a wide class of controlled phenomena under optimization.
Initially, the landscapes of linear time invariant systems are analyzed and compared with the quantum system case, then some analysis and a conjecture are offered on the fully non-linear case \ref{feqn}.
We simply remark that the landscape analysis for chemical and material science and for evolutionary biology follow a somewhat different track due to their complexity and the general nature of being open to the environment.
See the sufficient conditions in table \ref{tabsum}, and details in \cite{optichemeasy, optevo}.

There is large freedom to choose a cost function $J$ for both LTI (see \ref{lti}) and non-linear systems.
For control systems, both LTI and non-linear, which evolve on $\mathbb{R}^N$, a common and favorable choice is:
\begin{align}
	J(X) = ||\vec{X}(T) - \vec{X}_{G}||^2
\end{align}
where $\vec{X}(T)$ is the finial time state and $\vec{X}_{G}$ is the goal (to be minimized in this case).
This function is straightforwardly confirmed to possess no saddles or local optima by a gradient calculation.
For systems which evolve on more general state spaces than $\mathbb{R}^N$, other cost functions are appropriate and a fully general canonical choice is not widely agreed upon.
However, the conclusions of this work apply to any cost function $J$ which possess no local optima \emph{of its own} as a function of the state (rather than the control), even in the fully non-linear case.

\subsection{Autonomous linear systems with unrestricted controls are almost all trap free}

The study of linear control systems is pervasive in theory \cite{Antolin, kwaker} and applications \cite{Driels}, which motivates the study of their landscapes, as does the potential for using them as a stepping stone to analyzing the landscape of the fully non-linear case \ref{feqn}.

Linear, time invariant (LTI) systems are defined by the equation:
\begin{align}
\label{lti}
\frac{d \vec{X}}{dt} = A\vec{X}(t) + w(t) \vec{b}
\end{align}
where $\vec{X}(t) \in \mathbb{R}^N$ is the state vector, $A$ is an $N \times N$ matrix, $\vec{b} \in \mathbb{R}^N$ is a vector coupling to the control.
Equation \ref{lti} warrants explicit comparison to \ref{scheqn} which is a \emph{bi-linear} control system as a product of the state and the control is present.

For system \ref{lti} it is possible to check the three assumptions of landscape analysis directly, showing that they almost always hold in the space of all systems of form \ref{lti}.

\subsubsection{Assumption I: the system is controllable}

The first assumption of controlability has a clear analogue to that of assumption I in quantum control, i.e. all states $\vec{X}$ can obtained by applying some control $w(t)$.
A sufficient condition for this to hold is known.
One must check that the \emph{controllability matrix} is full rank \cite{kwaker}, particularly that the columns of
\begin{align}
	\label{conmat}
	\left[ \vec{b}, A \vec{b}, \ldots , A^{(N-1)} \vec{b} \right]
\end{align}
form a basis, or equivalently that $\vec{b}$ is a cyclic vector of $A$.
One can readily check that this property holds for typical $A$ and $\vec{b}$ (see Appendix \ref{LTIapp1}) in a similar sense to the quantum case.
An analogous condition is developed for the linear time varying case in \cite{tvcon}.

\subsubsection{Assumption II: The system is locally controllable in the direction of the cost function gradient}

LTI systems uniquely have the property that controllability is equivalent to local controlability (see Appendix \ref{LTIapp2}), and thus it follows that assumption II, holds whenever assumption I holds.

\subsubsection{Assumption III: The control resources are unrestricted}

In the LTI context, this assumption remains unchanged compared with the quantum control case.

\subsection{Control of non-linear systems}

\label{bigf}

In this section, landscapes for systems of the form \ref{feqn} will be discussed.
See also the paper by A. Fradkov which introduces this special issue on Norbert Weiner.

We first note that the Schr\"odinger equation, and the LTI equation are special cases of \ref{feqn}, as well as a Schr\"odinger equation non-linear in the control \cite{mepol, turc}.
The primary open question is that of the fullest possible scope of the landscape observations in this paper.
In order to tentatively assess this topic, we investigate the same three assumptions in this broader context of the control of non-linear systems.

We see that the first assumption of global controllability can be assessed in a wide variety of cases and that it is known to hold for a large and practically relevant class of non-linear control systems.
We further argue that the third assumption is on an identical footing to the LTI and quantum control cases.
The conclusion of this section is that gaining a deeper understanding of the second assumption is central to a fuller analysis of the landscapes of non-linear systems in general.

In the quantum case it was shown that \emph{almost all} (in a specific sense that the failing set is null) quantum control systems are locally controllable.
We argue that the same mathematical tool set as that used in quantum mechanics \cite{mesing} can be deployed in the assessment of the status of the second assumption in the non-linear case, but we do not give a full proof that the analogous conclusion holds in the non linear case.
We highlight that assessing the status of the second assumption in the non-linear case is the key open issue in understanding their control landscapes.

\subsubsection{Assumption I: The system is controllable}

Criteria for the controllability for several large classes non-linear systems are known \cite{lukes1972global, sus, Hirschnlcon}.
Specifically and importantly, the closest generalization of the controlability conditions applied to quantum (bilinear) and LTI control systems are developed in \cite{Sontag, lukes1972global}, wherein it is shown that a sufficiently small non-linear term added to an LTI system will not violate global controllabilty.
This circumstance allows the conclusons about assumption one, that almost all LTI and quantum systems are globally controllable, to be pushed directly through to non-linear systems.

In \cite{Sontag} several `rank' conditions are described, which are analogous to the results in \cite{altifini} for quantum systems and to \cite{zab} for LTI systems.
In \cite{lukes1972global}, it is shown that if a system has a controllable linear part and the non-linear part is `small enough' in a specific and unrestrictive sense, then the overall system is also globally controllable.
It follows from the observation that almost all LTI systems are controllable and the conclusions of \cite{lukes1972global} that, a very rich class of non-linear control systems are globally controllable.
In this sense, understanding the linear case serves as a gateway to understanding which of the non-linear family of systems are globally controllable, and to constructing a multitude of globally controllable examples.

\subsubsection{Assumption II: The system is locally controllable in the direction of the cost function gradient}

One measure of the \emph{complexity} of a control system is its degree of local controlability (i.e., the number of linearly independent directions the state can be steered in by infinitesimally varying the control).
These quantities can be expressed as:
\begin{align}
	\left\{ \frac{\partial \vec{X}}{\partial w_k} \right\}
\end{align}
for each control variable $w_k$ or, if the control is a function of time as the functional derivative:
\begin{align}
	\label{nlrank}
	\left\{ \frac{\delta \vec{X}}{\delta w(t)} \right\}
\end{align}assumption II
If the span of the set of \emph{all} such derivatives contains $\nabla J$ for every control, then it is possible to infinitesimally vary the control in a way which increases $J$ everywhere on the system's landscape.
The gradient of fidelity with respect to the control is then non-zero, and gradient ascent can continue to increase fidelity.
The presence of more variables (i.e., nominally, additional system complexity) can aid the simplicity of optimal control by better assuring satisfaction of assumption II.
For important work offering conditions on the local controlability of non-linear systems see \cite{kunze} and for an assessment of their local controlability specifically at equilibrium points see \cite{susnlc}.

\subsubsection{Assumption III: sufficient resources}

The overall nature of the required resources in the control of non-linear systems is parallel to that in LTI and quantum control.
In the case of unconstrained resources, the argument that the control landscape will be trap free (given the satisfaction of other two assumptions) is identical to the quantum and LTI cases (Appendix \ref{trapapp}).
Further work is needed to investigate the effect of non-linearity on which resource constraints introduce landscape traps, as this remains an open issue.

Importantly, satisfaction of the three assumptions for a non-linear system is sufficient to ensure a trap free landscape.
However, we at this time cannot assess what is \emph{typical} for non-linear systems in the same sense as the quantum or LTI cases.
The fullest range with which the three assumptions hold remains open for investigation and is a topic of prime importance.

\section{Conclusions and Future Work}

We have reviewed numerical and experimental evidence motivating a unifying mathematical framework for understanding control landscapes.
We presented the state of the field of quantum control landscapes, and explained the generalization of these ideas to the control landscapes of more general systems.
It has further been argued that an explanation for this commonality may be found in the control landscapes of general non-linear control systems.
It is very interesting that of system complexity should be conducive the simplicity of the landscape critical point topology, as singular critical points are known to be very unlikely to exist in this scenario.
Furthermore, sections \ref{gener} and \ref{beyo} clearly indicate that the same set of primary assumptions apply to establishing control landscape topology to a vast expanse of systems and domains.

It is interesting to note that within the optimal control problems discussed herein, a function of the following form is being optimized.
\begin{align}
	\label{funform}
	F(w) := J(x_T(w))
\end{align}
The `no free lunch' theorem \cite{nofreelunch,nofreelunchsearch} states that, averaged over all functions to optimize, no one algorithm out performs any other.
It is notable that this result isn't an issue in control, as only a special functional form is being optimized, this observation is consistent with the results and observations in this work, and we further speculate the existence of a free lunch upon seeking optimal control.

There are many additional directions to explore in the domain of general system control landscapes analysis.
These include systems having time dependent drift dynamics and stochastic systems.
For the systems analyzed in this paper, all time dependence enters the systems through the control.

Based on the experimental, analytic and numerical evidence discussed, the authors conjecture that systems of the form \ref{feqn} are almost all trap free and that their convergence rate to a globally optimal control can be shown to imply that only a small fraction of the control space needs to be explored to discover such an optimal control.
Yet, the fullest extent to which there is a free lunch in control remains to be resolved and stands as a challenge to assess.

\vskip6pt

Benjamin Russell thanks the NSF (grant no. CHE-1464569) and Herschel Rabitz thanks the Templeton foundation (grant no. 52265) for funding

The authors would like to thank Shanon Vuglar and Robert Kosut for helpful discussions.


\bibliographystyle{unsrt}
\bibliography{mybib}

\appendix

\section{Proof that assumptions are sufficient for a trap free landscape}
\label{trapapp}

Given a control system of the form \ref{feqn}, here an informal proof is given that the three assumptions are sufficient for trap free landscape.
Firstly, we assume that the cost function $J$ is trap free itself, that the manifold of states is $M$ and the control space is $C$.

We will denote the end-point map by $V_T$, which maps a control $w$ to the corresponding solution to $\ref{feqn}$, and the overall fidelity of a control $w$ to be $\mathcal{F}[w] = J(V_T[w])$.
The overall derivative of fidelity with respect to the control can be obtained by the functional derivative chain rule:
\begin{align}
\frac{\delta \mathcal{F}}{\delta w} = \frac{\delta \mathcal{J}}{\delta x_T} \circ \frac{\delta x_T}{\delta w} \label{chain}
\end{align}
Here the three assumptions are rephrased in geometric terms.
In these terms assumption I, which is controlability, is that $V_T$ is a surjective function $V_T: C \rightarrow M$ when all controls are permitted, which is itself exactly assumption III.
Assumption II, $V_T$ which is local controlability in the direction $\nabla J \big|_{x_T}$, is that the the image of the derivative (push forward) $\delta V_T \big|_{x_T}$ is not orthogonal to $\nabla J \big|_{x_T}$ for all controls $w \in C$.
The three assumptions together, yield two geometric properties of $V_T$ and subsequently of $\mathcal{F}$.
Firstly, $\forall g \in M$, $\exists w \in C$ s.t. $V_T[w]=g$.
Secondly, there does not exist a control $w \in C$ s.t. $\langle dV_T \big|_{x_T} [\delta w], \nabla J \big|_{x_T} \rangle=0$.

Examining equation (\ref{chain}), one see that there are two types of critical points $\frac{\delta \mathcal{F}}{\delta w} = 0$.
The case $\frac{\delta J}{\delta x_T} = 0$ is excluded by assumption that $J$ does not possess traps of its own.
The second case $\frac{\delta x_T}{\delta w} = 0$, is excluded by assumption II, that $\langle dV_T \big|_{x_T} [\delta w], \nabla J \big|_{x_T} \rangle=0$.

The assumptions I and II, means that the image of $V_T$ is the full space of states.
The assumption II $\langle dV_T \big|_{x_T} [\delta w], \nabla J \big|_{x_T} \rangle=0$, implies that the image of $\frac{\delta x_T}{\delta w}$ is not orthogonal to $\nabla J \big|_{x_T}$.
Together these two statements imply that gradient ascent can continue until the objective is maximized, as away from the global maxima and minima of $\mathcal{F}$ the gradient is not zero, and the objective can be maximized dues to controlability.
Together these statements imply that gradient ascent can maximize $J$, and that the landscape is free from traps.

We further note that if $J$ is known to possess saddles (as is the case in many quantum control \cite{dom1}), then these only translate into saddles on the control landscape rather than true traps.

\section{LTI systems are almost all controllable}

\label{LTIapp1}
One can readily check that, over all $A, \vec{b}$ the probability of selecting a non-controllable system is zero or, equivalently, that the set of $A, \vec{b}$ which correspond to uncontrollable systems form a null set.
First notice that the probability of selecting diagonalizable $A$ is one when $A$ is chosen at random.
Thus, it will have distinct, non-zero eigenvalues from which it follows:
\begin{align}
\left[ \vec{b}, A \vec{b}, A^{2} \vec{b}, \ldots, A^N \vec{b} \right]
=
Q \left[ \vec{c}, D \vec{c}, D^{2} \vec{c}, \ldots, D^N \vec{c} \right] \nonumber
\end{align}
where $\vec{c} = Q^{-1}\vec{b}$.
Now observe that the determinant of the controlability matrix \ref{conmat} is now equal to the determinant of the Vandermonde matrix generated by the eigenvalues of $A$ ($a_n$ say) and the matrix $diag(\vec{c})$, i.e. $\det(V) \text{diag}(\vec{c}) = c_1 \ldots c_N \det(V)$.
This determinant is clearly non-zero unless any $a_n$ is zero, which itself happens with probability zero.

\section{Equivalence of assumptions 1 and 2 for LTI systems}
\label{LTIapp2}

The variation of the endpoint map can be found  directly from the LTI defining equation to also be an LTI equation.
\begin{align}
\delta X(t) = A (\delta X)(t) + (\delta w)(t) \vec{b} \nonumber
\end{align}

We also note that the end point map for an LTI system, unlike almost all non-linear systems, can be found in closed form:
\begin{align}
X(T) = C e^{t A} X(0) + \int_{0}^{T} w(t) e^{(T-t)A} \vec{b} dt \nonumber
\end{align}
See any text book on LTI control systems for this derivation.

\end{document}